\documentclass[traditabstract]{aa}
\usepackage{xcolor}

\usepackage[T1]{fontenc}
\usepackage[utf8]{inputenc}
\usepackage{ae,aecompl}
\usepackage{lipsum}
\usepackage{pdflscape}

\usepackage{graphicx}	
\usepackage{amsmath}	
\usepackage{amssymb}	
\usepackage{url}
\usepackage[colorinlistoftodos,prependcaption,textsize=tiny]{todonotes}
\usepackage[font=footnotesize]{subfig}
\usepackage{soul}
\usepackage{nccmath}
\usepackage{pdflscape}	%
\usepackage{multicol}
\usepackage{enumerate}
\usepackage{times,epsfig,natbib,amssymb,amsmath,graphics,longtable,lscape,threeparttable}

\begin{document}
\title{The effects of varying colour-luminosity relations on type Ia supernova science}


\author{
S. Gonz\'alez-Gait\'an\inst{1}\thanks{E-mail: gongsale@gmail.com}
\and T. de Jaeger\inst{2,3}
\and L. Galbany\inst{4}
\and A. Mour\~ao\inst{1}
\and A. Paulino-Afonso\inst{1}
\and A.~V. Filippenko\inst{3,5}
}
\institute{
CENTRA, Instituto Superior T\'ecnico, Universidade de Lisboa, Av. Rovisco Pais 1, 1049-001 Lisboa, Portugal.
\and Institute for Astronomy, University of Hawaii, 2680 Woodlawn Drive, Honolulu, HI 96822, USA.
\and 
Department of Astronomy, University of California, Berkeley, CA 94720-3411, USA.
\and Institute of Space Sciences (ICE, CSIC), Campus UAB, Carrer de Can Magrans, s/n, E-08193 Barcelona, Spain.
\and 
Miller Institute for Basic Research in Science, University of California, Berkeley, CA 94720 USA.\\
}

\abstract{

The success of Type Ia supernova (SN~Ia) distance standardisation for cosmology relies on a single global linear relationship between their peak luminosity and colour, the $\beta$ parameter. However, there are several pieces of evidence and physical reasons to believe that this relation is not universal and may change within different subgroups, or even among individual objects. In this work, we allow $\beta$ to vary among subpopulations with different observed properties in the cosmological fits. Although the inferred cosmological parameters are consistent with previous studies that assume a single colour-luminosity relation, we find that the SN data favour nonuniversal distributions of $\beta$ when split according to SN colour and/or host-galaxy mass. For galaxy mass, we obtain a $\beta$-step relation in which low $\beta$ values occur in more massive galaxies, a trend that can be explained by differing dust reddening laws for two types of environments. For colour, we find that bluer/redder SNe~Ia are consistent with a lower/larger $\beta$. This trend is explained with $\beta$ being a combination of a low intrinsic colour-luminosity relation dominant in bluer SNe and a higher extrinsic reddening relation dominant at redder colours. The host galaxy mass-step correction always provides better distance calibration, regardless of the multiple $\beta$ approaches, and we suggest that it may come from a difference in intrinsic colour-luminosity properties of SNe~Ia in two types of environments. Additionally, we find that blue SNe in low-mass environments are better standard candles than the others.
}

\keywords{supernovae:general, cosmology:observations}
\authorrunning{Gonz\'alez-Gait\'an et al.}
\titlerunning{Effects of $\beta$ variation on SN science}
\maketitle



\section{Introduction}
After sustained confirmation of the accelerating expansion of the
Universe over the past two decades \citep[e.g.,][]{DES19,Jones19,Scolnic18,Betoule14,Perlmutter99,Riess98}, Type
Ia supernova (hereafter SN~Ia) cosmology has reached a mature state in which systematic
uncertainties are on the order of statistical uncertainties \citep{Brout19,Conley11}. Different aspects of SN observations, SN populations across redshifts, and SN standardisation, among other factors, require careful scrutiny to better constrain the equation-of-state parameter $w = P/(\rho c^2)$ of the unknown dark energy. Large
efforts are being dedicated to understand and diminish calibration effects by using well-understood photometric systems across a range of redshifts \citep{Brout19,Betoule13}, by reducing selection biases and contamination \citep{Kessler17}, by improving light-curve and cosmological fitters \citep[e.g.,][]{Hinton19,Mandel17,Burns14}, by including further observables to standardise SNe \citep{Bailey09}, among others.

SNe~Ia are not perfect standard candles, yet empirical relations between their observed peak luminosity and colour \citep[bluer SNe~Ia are more luminous;][]{Tripp98}, as well as between peak luminosity and light-curve width \citep[SNe~Ia with wider light curves are more luminous;][]{Phillips93}, make them good distance indicators to $\sim3$--5\% precision \citep[e.g.,][]{DES19}. Moreover, it has been shown that an additional luminosity correction related to host-galaxy stellar mass can further increase the precision of SN~Ia distances: SNe~Ia in more massive galaxies are more luminous after other corrections \citep{Sullivan10,Lampeitl10,Kelly10}. This host-galaxy mass correction is possibly a proxy for a difference in stellar-population age of the progenitors \citep{Kang20,Rigault18}.

While some studies have indicated that colour is the main driver of SN variability \citep[][although see \citealt{Leget19,He18}]{Sasdelli15,Kim13}, the origin of the colour-luminosity relation is still under heavy debate.

On the one hand, the effects of dust along the line of sight are expected to explain it, at least partly, by making the observed SN magnitude dimmer through light absorption, and the colour redder through selective scattering and absorption of shorter wavelengths. This interplay depends on the dust properties and is captured in the reddening law, which is usually parameterised in the optical through the total-to-selective extinction, $R_V=A_V/(A_B-A_V)$. It has been demonstrated that the reddening law is not a universal constant --- dust properties vary, as shown in our own Milky Way (MW; \citealt{Nataf16}) and toward other galaxies \citep{Salim18,Cikota16}. Even some individual well-observed SNe are known to have very peculiar $R_V$ values, inconsistent with the average MW value of 3.1 \citep{Gutierrez16,Foley14}. This means that dust properties may vary from one to another SN line of sight, and assuming a common reddening law for all SNe is an oversimplification that may affect the calibration of SNe~Ia. This is particularly true if there is an evolution of dust properties with redshift or a dependence of the reddening law with the SN properties used in the calibration. Hints for dependence of extinction and/or $R_V$ on intrinsic SN properties do exist \citep{Forster13,Foley11b}.

On the other hand, to further complicate the situation, SNe~Ia are known to be intrinsically different, with possible evidence for numerous progenitor and/or explosion scenarios (see \citealt{Maoz14} for a review). The variation of ionisation state and temperature with phase, which relate to the colour, opacity, and brightness evolution of each SN, make less-luminous and cooler SNe~Ia also redder owing to an earlier onset of the recombination of iron-peak elements \citep{Kasen09}. Although theoretically a large portion of the intrinsic luminosity and colour variation may be explained through a single parameter, some unaccounted colour variation is clearly present \citep{Hoeflich17}, and multiple intrinsic colour-luminosity relations may be relevant \citep{Mandel14}, casting clear doubt on a common intrinsic colour-luminosity relation for all normal SNe~Ia.

The colour-luminosity relation of SNe~Ia is thus probably coming from a combination of intrinsic and extrinsic factors that cannot easily be modelled with a simple universal linear relation, as has traditionally been done. This was recognised early \citep{Conley07}, since the $R_V$ inferred from the colour-luminosity relation of cosmological fits is always intriguingly lower than the MW value. 

More recently, \citet{Mandel17} have addressed this issue by going beyond the traditional linear regression of colour-magnitude to include both an intrinsic colour-luminosity relation and an external host-galaxy dust extinction component. By assuming a single $R_V$ parameter for all SNe~Ia, they report a consistent value with the MW. Furthermore, \citet{Brout20} use a colour model that includes intrinsic and dust effects, but this time allowing a distribution of $R_V$ values. With the help of extensive survey simulations, they find a relation between $R_V$ and host-galaxy mass which naturally explains the host mass-step commonly used to further reduce the Hubble residuals. Their result has been recently supported by the study of \citet{Johansson21} with $R_V$ values obtained individually through optical/NIR data without the need of a mass step. On the other hand, other recent studies contest the riddance of this mass-step: \citet{Thorp21} through hierarchical Bayesian fits and \citet{Uddin20,Ponder20} through extinction-reduced near-infrared Hubble residuals that still correlate strongly with host galaxy mass.

In this work, we investigate the question of varying colour-luminosity relations with a simple approach based directly on the cosmological sample of SNe~Ia and inspired by the traditional linear regression used in cosmological studies (see Section~\ref{txt:montepython}). By building several realistic distributions of individual $R_V$ values for each SN in the dataset, we compare the cosmological results with a standard universal $R_V$ approach in Section~\ref{txt:fixed_beta}. In Section~\ref{txt:free_beta}, we investigate the cosmology when allowing several multivariate colour-luminosity relations to be free in the fit, an inexpensive approach that can be readily used in routine SN cosmological analyses with enlightening results. We gauge for systematic biases in the fitting in Section~\ref{sec:like}.  We conclude in Section~\ref{sec:conc}.

\section{Methodology}\label{txt:montepython} 

The classical well-known \citet{Tripp98} formula for SN~Ia standardisation uses a simple linear calibration of observed magnitude, $m_B$, with observed colour, $\mathcal{C}$, and light-curve width, $X$, to obtain a corrected $m_B^{\mathrm{corr}}$,

\begin{ceqn}
\begin{align}
m_B^{\mathrm{corr}} = m_B + \beta \mathcal{C} - \alpha X. 
\label{eq_standardisation}
\end{align}
\end{ceqn}
\noindent
To infer cosmology, one compares this to a model magnitude, $m_B^{\mathrm{mod}}$, dependent on the redshift, $z$, and the cosmological model (here represented with the matter density, $\Omega_m$, and the dark energy equation-of-state parameter, $w$) through the luminosity distance $D_L(z,{\rm H}_0,\Omega_m,w)$ in Mpc:

\begin{ceqn}
\begin{align}
m_B^{\mathrm{mod}} =  M_B + 5\log_{10}D_L(z,{\rm H}_0,\Omega_m,w) + 25.
\label{eq_model}
\end{align}
\end{ceqn}
\noindent
Here $M_B$, $\alpha$, and $\beta$ are general nuisance parameters found in the cosmological fitting process. The constants $\beta$ and $\alpha$ contain respectively the colour-luminosity and width-luminosity relations which are thus assumed universal and linear. For the case of the colour-luminosity relation, this would be expected if it arises uniquely from dust with the same properties, making extinguished SNe simultaneously redder and fainter in the same way. In this paper, we loosen this constraint and investigate the effects on cosmology of differing colour-magnitude relations --- that is, the effect of having multiple $\beta$ values for different populations of SNe. If $\beta$ were entirely due to dust, its relation with the reddening law would be $\beta = R_B = R_V+1$.

Ideally, a more complete way to properly take into account both intrinsic and dust reddening relations with luminosity would have an intrinsic SN colour $c_{\mathrm{int}}$ related to luminosity through an intrinsic $\beta_{\mathrm{int}}$ relation, and a dust component expressed through the colour excess $E(B-V)=A_B-A_V$ attenuating the $B$-band magnitude by $(R_V+1)E(B-V)$. The total colour is thus $\mathcal{C}=c_{\mathrm{int}}+E(B-V)$ and the \citet{Tripp98} equation becomes
 
\begin{ceqn}
\begin{align}
m_B^{\mathrm{corr}} = m_B+\beta_{\mathrm{int}}c_{\mathrm{int}} + (R_V+1)E(B-V) - \alpha X .
\label{eq_standardisation_int}
\end{align}
\end{ceqn}
\noindent
 Such a comprehensive treatment, however, requires proper calculation of the colour excess and intrinsic colour for each SN, a difficult task that demands accurate techniques and multiwavelength data for SNe, even at high redshift. This has not been accomplished yet for any cosmological SN sample.
 
 Furthermore, it is important to note that additional terms can refine this calibration, in particular the host stellar mass-step function now routinely used in cosmological fits. This is captured by a mass-step term, $\delta_{M}$, added to the previous equations depending on the host mass: 
 \begin{ceqn}
\begin{align}
\delta_M = \begin{cases}
0 &\text{if $M_{\mathrm{stellar}}<10^{10}\,M_{\sun}$},\\
\delta_M &\text{if $M_{\mathrm{stellar}}>10^{10}\,M_{\sun}$}.
\end{cases}
 \label{eq_massstep}
 \end{align}
 \end{ceqn}
 
 Throughout this paper, we minimise a likelihood that combines SNe~Ia with constraints from the cosmic microwave background (CMB) data. As observational data, we use the ``Joint Light-curve Analysis'' (JLA) sample released by \citet{Betoule14}, which consists of 740 SNe~Ia spanning a large range in redshift (up to $z \approx 1.3$) and the CMB data from \citet{planck18}. The likelihood function used for the SNe~Ia is defined as

\begin{ceqn}
\begin{align}
-2\,\mathrm{ln}(\mathcal{L}) = \chi^2 = \left  (\vec{m}_{B}^{\mathrm{corr}} - \vec{m}_{B}^{\mathrm{mod}} \right)^T  \mathsf{C}^{-1}   \left (\vec{m}_{B}^{\mathrm{corr}} - \vec{m}_{B}^{\mathrm{mod}} \right),
\label{eq:likelihood}
\end{align}
\end{ceqn}
where $\vec{\mu} = \vec{m}_B^{\mathrm{corr}} - \vec{m}_{B}^{\mathrm{mod}}$ is the vector of Hubble residuals for all 740 SNe~Ia, and $\mathsf{C}$ is the covariance matrix, whose diagonal terms are
\begin{ceqn}
\begin{align}
\begin{split}
\sigma^{2}_\mathrm{tot}  = \;&\sigma^{2}_{m_{B}} + (\beta \sigma_{C})^2 + (\alpha \sigma_{X})^2 \\
&-2\beta\mathsf{C}_{m_B,\mathcal{C}}+ 2\alpha\mathsf{C}_{m_B,X}-2\alpha\beta\mathsf{C}_{X,\mathcal{C}}\\
&+(\sigma_\mu^z)^2+(\sigma_{\mathrm{lens}})^2+\sigma^{2}_\mathrm{int}.
\end{split}
\label{eq:sigma}
\end{align}
\end{ceqn}
Here, the first three terms are the errors on the fitted light-curve parameters ($m_B$, $\mathcal{C}$, and $X$), the second three terms are the covariances between them, and the last three terms correspond to the error arising from redshift uncertainties ($\sigma_{\mu}^z$), the uncertainty caused by gravitational lensing ($\sigma_{\mathrm{lens}}$; \citealt{Jonsson10}), and the intrinsic variation ($\sigma_\mathrm{int}$) per SN sample or SN population. The intrinsic variation (together with the cosmological and nuisance parameters) is obtained in an iterative manner by ensuring that the $\chi^2$ per degree of freedom for a given SN sample is close to 1 or, alternatively, by minimising a restricted log-likelihood as by \citet{Betoule14}.

Furthermore, in  Eq.~\ref{eq:likelihood} we add to the Hubble residuals the redshift-dependent bias corrections, $\Delta\mu_{\mathrm{bias}}(z)$, from \citet{Betoule14}. These arise mainly from selection effects and are calculated through large survey simulations. It is important to note that our ``JLA-type'' methodology consists of a simple fit based on the likelihood of Equation~\ref{eq:likelihood} and a bias correction dependent only on redshift. This likelihood has been extensively used in the literature; however, studies show that these assumptions may introduce some biases in the cosmological and nuisance parameters \citep{Kessler17}. We study and discuss the impact of the chosen likelihood in Section~\ref{sec:like}.

Finally, we use the constraints of the background level from the CMB \citep{planck18} summarised by \citet{Betoule14} with Gaussian priors on the baryon density $\Omega_{b} h^{2}$, the cold dark matter density $\Omega_{c} h^{2}$, and the approximated sound horizon angular size 100\,$\theta_{s}$. Our final prior has the same form as Equation~18 of \citet{Betoule14}, but the values used were updated to \citet{planck18}\footnote{Best fit from \url{https://github.com/brinckmann/montepython_public/blob/3.3/bestfit/base2018TTTEEE.bestfit .} \\Covariance matrix from \url{https://github.com/brinckmann/montepython_public/blob/3.3/covmat/base2018TTTEEE.covmat .}}.

To solve the global likelihood, we use the Python package \textsc{MontePython} developed by \citet{audren13} and \citet{brinckmann19}. \textsc{MontePython} uses Bayesian inference to estimate cosmological parameters. Prior knowledge of the parameters is used to obtain the full posterior probability distribution. For each Monte Carlo Markov Chain simulation, we run four chains of 50,000 steps each. To ensure convergence of our chains, we check that the Gelman-Rubin convergence ratio \citep{gelman92} is smaller than 0.05 for each parameter. If it is not the case, we increase the number of steps in each chain.

In order to compare the different cosmological fits, we calculate the Bayesian Information Criterion \citep[BIC;][]{Schwarz78}, defined as

\begin{ceqn}
\begin{align}
\mathrm{BIC} = k\,\ln{n}-2\,\ln{\mathcal{L_{\mathrm{max}}}} ,
\label{eq:BIC}
\end{align}
\end{ceqn}
where $k$ is the number of free parameters in the model, $n$ is the number of data points, and $\mathcal{L_{\mathrm{max}}}$ is the likelihood evaluated at the optimal parameters that maximise it. A given Model 1 is preferred over another Model 2 if $\Delta\mathrm{BIC} = \mathrm{BIC}_1 - \mathrm{BIC}_2 \gtrsim 2$ \citep{Kass95}. Note that besides taking into account the usual fit quality, the BIC penalises models with excessive free parameters. However, since the nuisance parameters $\alpha$ and $\beta$ are in the covariance (Eq.~\ref{eq:sigma}) and may thus bias the likelihood (see Sec.~\ref{sec:like}) and BIC, we also quote the Hubble residuals of each fit.

\section{Effect of assumed populations of $\beta$}\label{txt:fixed_beta}

\subsection{Mock $\beta$ distributions}\label{txt:sim}

In this section, we explore different scenarios of varying $\beta$ 
parameters assigned to each SN of the sample of
740 SNe~Ia. We follow a simple approach in which all observed variables of the SNe, like apparent magnitude, light-curve width, and colour are kept intact, while the $\beta$ parameter, normally a nuisance constant fitted in cosmology, is assigned a value (or distribution of values for each SN), and is kept fixed in the fit. We thus investigate the effect on the fitted cosmological parameters when having differences in the assumed $\beta$ values. We compare all of our scenarios to the standard case which has all SNe with the same fixed $\beta_0=3.02$, the value found when a single colour-luminosity relation was left free in the fit. In order to compare all these assumed scenarios in a consistent way, we always make sure that the final mean of the mock $\beta$ distribution is the same as in the original $\beta_0$ standard fit. This also means that the actual $\beta$ values of our simulations do not have a physical meaning but rather the differences between each SN or SN population are relevant. By doing this, we aim to find if the observed data have a preference for a given scenario with a $\beta$ distribution that is different from the standard one, and if the inferred cosmology is affected by such a choice. Naturally, with this approach we can only explore a very limited set of possible $\beta$ distributions; however, it may give us an idea of the effect of allowing different colour-luminosity relations in cosmological fits.

We use the standard \citet{Tripp98} calibration of Equation~\ref{eq_standardisation} (with mass-step), without separating the intrinsic and reddening contributions (Eq.~\ref{eq_standardisation_int}), since we do not yet have  good knowledge of the actual intrinsic colours and reddening estimates for this observed sample, and we want to avoid introducing any bias with simulations that are not fully understood. For each $\beta$ scenario, we follow the procedure highlighted in the previous section, minimising a combined likelihood of SNe~Ia and CMB to find the optimal cosmological model. We emphasise that for all scenarios considered here the free variables are always the same: $\alpha$, $M_B$, $\delta_M$, $\sigma_{\mathrm{int}}$, and the cosmological parameters (H$_0$, $\Omega_M$, $w$).

\begin{figure}
\centering
\includegraphics[trim=0.0cm 0.0cm 0.0cm 0.0cm,clip,width=1.0\linewidth]{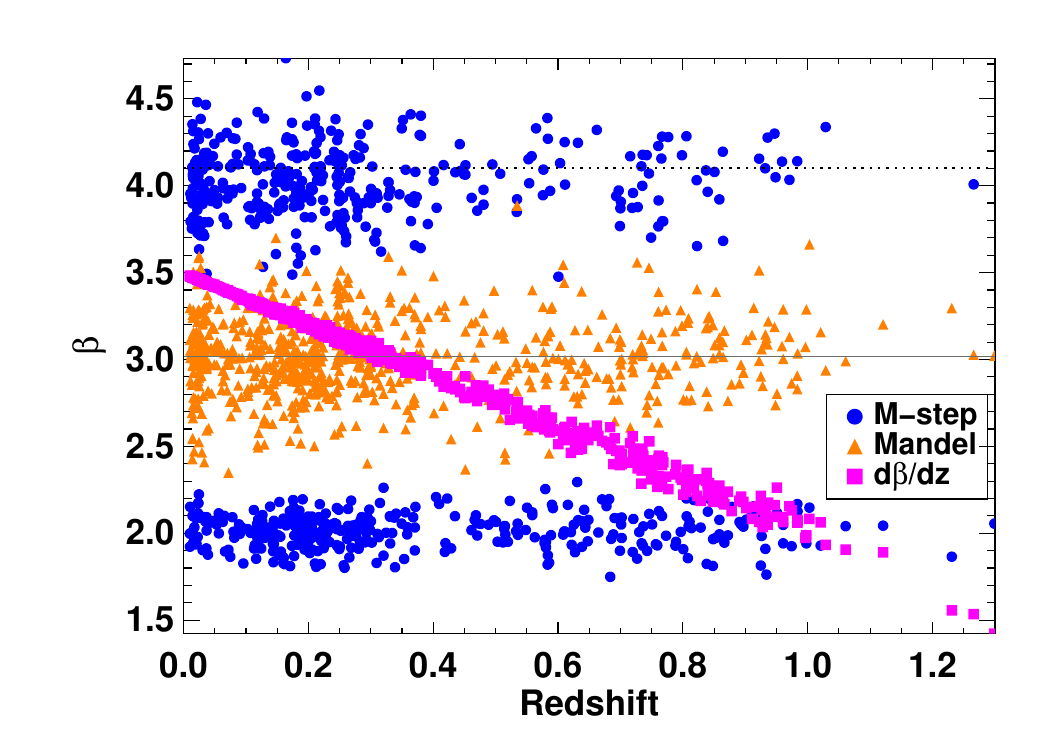}
\caption{Selected examples of simulated $\beta$ distributions based on (a) a bimodal population (with two different Gaussian widths) dependent on the host-galaxy mass (blue circles), (b) a 4-bin $R_V$ distribution (orange triangles) from \citet{Mandel11} based on observed colour, and (c) a continuous distribution dependent on redshift (purple squares). The grey solid line shows the $\beta_0$ of the standard 1 $\beta$ fit: all $\beta$ values of a given simulation are centred on this value --- that is, $<\beta_\mathrm{sim}>=\beta_0$. A horizontal dotted line shows the MW $R_V=3.1$ value.} 
\label{fig:beta-dist}
\end{figure}

We generate 232 mock $\beta$ distributions motivated by various observations of SNe~Ia and their environments (see three examples in Fig.~\ref{fig:beta-dist}). Below we explain each of them, going from simple bimodal models to individual $\beta_i$ for each SN.

\begin{enumerate}[(i)]
\item {\bf AB-model:} When modeling SN~Ia rates, a popular phenomenological model to account for the redshift evolution consists of two contributing $A+B$ populations \citep{Scannapieco05,Mannucci06}. The $A$-component is related to the old stellar population of evolved galaxies and contributes mostly to the rate at low redshift, whereas the $B$-component is associated with the young star-forming galaxies and contributes mostly at high redshift. With a bimodal SN~Ia rate calculated from the delay-time distribution of \citet{Perrett12} and a star-formation history of \citet{Li08}, we randomly assign all SNe within bins of $\Delta z=0.05$ to the $A$ or $B$ components, according to the expected rate at that redshift. It is known that the SNe in each component are also related to differences in environments, such as galaxy mass and star formation; for simplicity, we study those later independently (see the ``Mstep-model'' and ``X-model'' below) and we focus here on the rate evolution with redshift.

Once we choose the SNe belonging to each A or B component, we assign a $\beta$ value to each of the two populations. We explore differences between the colour-luminosity of the two populations in a range of $-2.0<\Delta\beta<2.0$ in bins of size $0.25$. We draw from Gaussian distributions centred at those $\beta$ values and explore two Gaussian widths for the two populations of $\sigma=0.05\,\beta$ or $\sigma=0.25\,\beta$. We end up with a total of 64 AB-model simulations. An example of a scenario with $\Delta\beta = \beta_B - \beta_A = -2.0$ would have all SNe of the A component (378 objects) in a Gaussian distribution of $\beta_A$ centred at 4.00 (with a $\sigma_A$ of 0.11 or 0.55), while the SNe of the B component (362 objects) would have $\beta_B$ centred at 2.00 (with a $\sigma_B$ of 0.02 or 0.10).

\item {\bf Mstep-model:} In recent years, there has been evidence for a residual dependence on host-galaxy mass ($M_{\mathrm{stellar}})$ after correction for light-curve width and colour leading to a third bimodal correction dependent on host-galaxy mass. We thus investigate the effect of those two populations in distinct environments having two different $\beta$ values by dividing the SNe into two groups according to the median host-mass value around $10^{10}\,M_{\sun}$, and varying $\beta$ in bins of size $0.25$ within $-2.0<\Delta\beta<2.0$. We try Gaussian distributions with two widths ($\sigma=0.05\,\beta$ or $\sigma=0.25\,\beta$) for a total of 64 Mstep-model combinations. An example scenario with $\Delta\beta = \beta_{M_1} - \beta_{M_2} = 2.0$ (see Fig.~\ref{fig:beta-dist}) has the SNe in low-mass galaxies (half of the sample) with a distribution of $\beta_{M_1}$ centred at 4.02 with a width of 1.00 and the SNe in the high-mass galaxies having $\beta_{M_2}$ centred at 2.02 with a width of 0.10.

\item {\bf X-model:} The light-curve width, here $X$ \citep{Guy07}, is strongly related to star formation in the environment and presumably progenitor age \citep[e.g.,][]{Hamuy95,Sullivan06b}. 
Therefore, we use the light-curve width as a proxy for age and investigate the cosmological effects of a bimodal distribution of $\beta$ changing with $X$. Again, we explore a difference between the two populations in the range $-2.0<\Delta\beta<2.0$ and with two Gaussian widths.

\item {\bf $R_V$-Mandel and Burns models:} By using multiwavelength (optical/near-infrared) photometry, \citet{Mandel11} and \citet{Burns14} obtain reddening laws along the line of sight of each SN~Ia, finding that they follow a relation of lower mean $R_V$ at increasing extinction. Both studies present their $R_V$ results in 4--5 bins of $A_V$ or $E(B-V)$ (see their Tables~3 and 6, respectively). To model this distribution with our data, we use the observed colour as an approximation to colour excess. We divide our SN sample in these bins and assign a reddening law $R_V$ to each from a Gaussian distribution with mean and sigma from the tabulated mean and standard deviation. Since the SN~Ia samples chosen for cosmology are biased to more ``normal'' SNe~Ia (they have colour cuts of $c \lesssim 0.2$--0.3), our distributions are restricted to only two of their bins. Additionally, we fit a power law of the $R_V(A_V)$ data of both studies to generate another $R_V$ simulation based on this fit. We note that the final simulated $\beta$ values are not directly equal to $R_V+1$ owing to our normalisation to the standard $\beta_0$ model. The Mandel model is shown in Figure~\ref{fig:beta-dist}.

\item {\bf $z$-model:} One of the most concerning systematics in SN cosmology stems from the evolution of observables and SN properties with cosmic time. There are indications that $\beta$ could change with redshift with a $d\beta/dz$ slope ranging from $-1.9$ to +1.0 \citep{Scolnic18,Conley11,Guy10,Kessler09}. We simulate such a scenario and assign a $\beta$ value to each SN according to its redshift by drawing from several $\beta$ distributions that follow slopes in the range $-2<d\beta/dz<2$ in steps of 0.25 with two Gaussian widths of $0.05\, d\beta/dz$ or $0.25\, d\beta/dz$ for a total of 32 $z$-simulations. The case of $d\beta/dz = -2.0$ with $\sigma=0.10$ is shown in purple in Figure~\ref{fig:beta-dist}.
\end{enumerate}

Each one of the 232 scenarios is simulated 100 times in a Monte Carlo that draws the parameters from the assumed distributions, as explained for each set of scenarios. For a given model, the resulting cosmological parameters and BIC diagnostics are the median of all 100 realisations. We note that the 3$\sigma$ dispersion of the BIC diagnostic for each scenario is always $< 0.15$, confirming that our choice of $\Delta$BIC $>2$ (see previous section) will correctly select statistically different scenarios (by about 6$\sigma$).

\subsection{Results}\label{txt:simres}
\begin{figure}
\centering
\includegraphics[trim=0.0cm 0.0cm 0.0cm 0.0cm,clip,width=\linewidth]{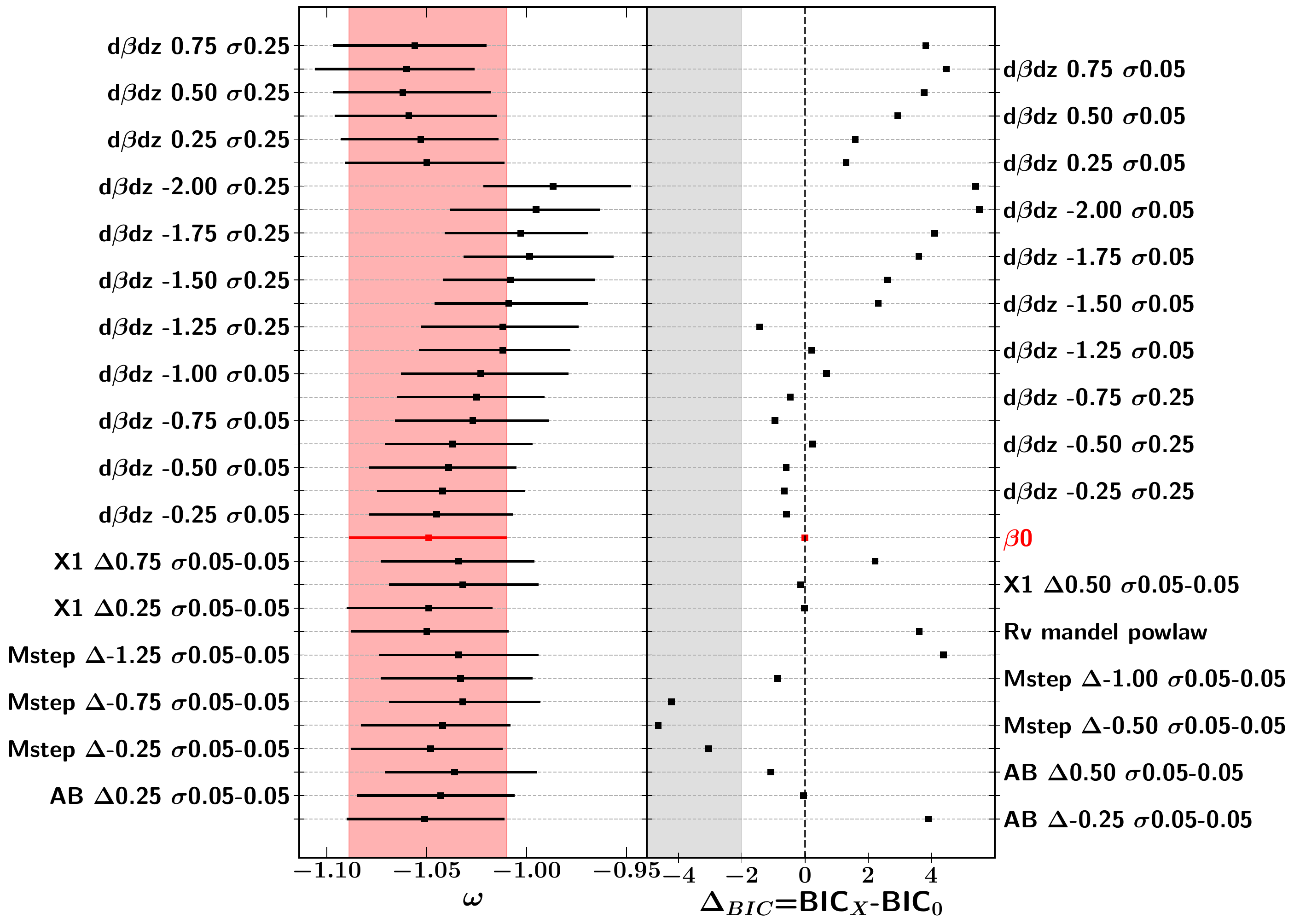}
\caption{\emph{Left}: Dark energy equation-of-state parameter $w$ for selected examples of mock $\beta$ distributions with their respective 1$\sigma$ uncertainties. The name of the distribution is on both sides of the plot, indicating if it is a $z$-model with positive or negative $d\beta/dz$ and the width of the distribution, an $X$-model, an Mstep model, or an AB-model with $\Delta\beta$ between both populations and their Gaussian width distribution (e.g., model ``$X$ $\Delta$--0.75 $\sigma$0.05--0.25'' is an X1-model with a $\beta$ difference of $-0.75$ between both populations and Gaussian widths of 0.05 and 0.25, respectively). The shaded area in red indicates the mean $w$ and 1$\sigma$ for the standard $\beta_0$ case. All models are consistent within the uncertainties. \emph{Right}: Difference between the BIC of the simulated fit and the standard $\beta_0$ fit for the same distributions as in the left. $\Delta\mathrm{BIC}<-2$ indicates a clear preference for the simulated model and is shaded in grey.}
\label{fig:simres}
\end{figure}

One of the main goals of the generation of mock $\beta$ distributions is to gauge the effect of varying colour-luminosity relations in the estimation of the cosmological parameters. We show the changes in the dark energy equation-of-state parameter for a subset of simulated distributions in the left panel of Figure~\ref{fig:simres}. These represent the best 35 scenarios according to the BIC diagnostic. As can be seen, there are net variations of at most $\Delta w\lesssim0.05$. However, taking into account the fit uncertainties, these differences are always $<1\sigma$, completely consistent with the standard single-$\beta$ approach. We can conclude that for the current dataset, the different simple scenarios investigated do not significantly affect the cosmological results obtained in the past. However, with the more-accurate observations and reduced uncertainties of coming surveys, the situation could change in the near future.

On the other hand, there is clear evidence that the data favour some of the simulated scenarios. By looking into the fit quality with the BIC diagnostic (see Eq.~\ref{eq:BIC}) applied to all mock scenarios and compared to the standard case, we are in a position to evaluate if some $\beta$ distributions are preferred over a single-value case. We note that all fits have the same number of free parameters and number of data points, as well as the same average $\beta$, so that the comparison is even more straightforward and does not suffer from likelihood biases (see Sec.~\ref{sec:like}). In the right panel of Figure~\ref{fig:simres}, we show the difference in BIC of the selected sample of simulated distributions compared to the standard single $\beta_0$ case. Although several scenarios have BIC values lower than the standard case ($\Delta$BIC $ < 0$), only bimodal $\beta$ distributions separated by host-galaxy mass have significantly better fits than the single-$\beta$ case ($\Delta$BIC $ < -2$). These correspond to differences of $\Delta\beta = 0.25$--1.00 between two SN populations in host galaxies with different masses. SNe in host galaxies with larger mass are consistent with $\beta$ values that are smaller than for SNe~Ia in lower-mass galaxies. This result is qualitatively consistent with the observations of \citet{Salim18} for a large sample of galaxies and their inferred masses and reddening laws when including quiescent galaxies), and also with the findings of \citet{Brout20} for SNe~Ia. 

The fact that we find scenarios with significantly better fits than the standard single $\beta$ case demonstrates that, as expected, a single universal value of $\beta$ is a limiting simplification in cosmology. We emphasise that we have only explored here a small set of hypothetical possibilities within an infinite set of $\beta$ distributions, and evidently there should be many better fits as we choose scenarios that approach reality more closely. This will be further studied in the next section when we allow multiple $\beta$ parameters to be free in the cosmological fits.

\section{Effect of varying $\beta$ within subgroups in cosmological fits}\label{txt:free_beta} 

In this section, instead of fixing $\beta$ values for each SN based on simulated distributions, we perform cosmological fits to the SN sample allowing several free $\beta$ parameters to be directly inferred from the fit. Since it is impossible at present to derive an individual $R_V$ reddening law for every SN, and even less an individual intrinsic colour-luminosity relation, 
we allow $\beta$ to vary for a subsample of objects based on different observed parameters such as colour, host-galaxy stellar mass, light-curve width, redshift, and $B$-band magnitude at maximum brightness. For each observable, the whole sample is split into 2, 3, or 4 bins (divided according either to the number of SNe in each bin or alternatively to the bin size); then we fit Equation~\ref{eq_standardisation} (with mass-step) allowing 2, 3, or 4 $\beta$ values and compare to 1 $\beta$, the standard case similar to \citet{Betoule14}. The same fitting procedure described in Section~\ref{txt:montepython} is used --- that is, we minimise a likelihood that combines SNe~Ia with constraints from the CMB.

\subsection{Cosmological parameters}

\begin{figure}
\centering
\includegraphics[width=1.0\columnwidth]{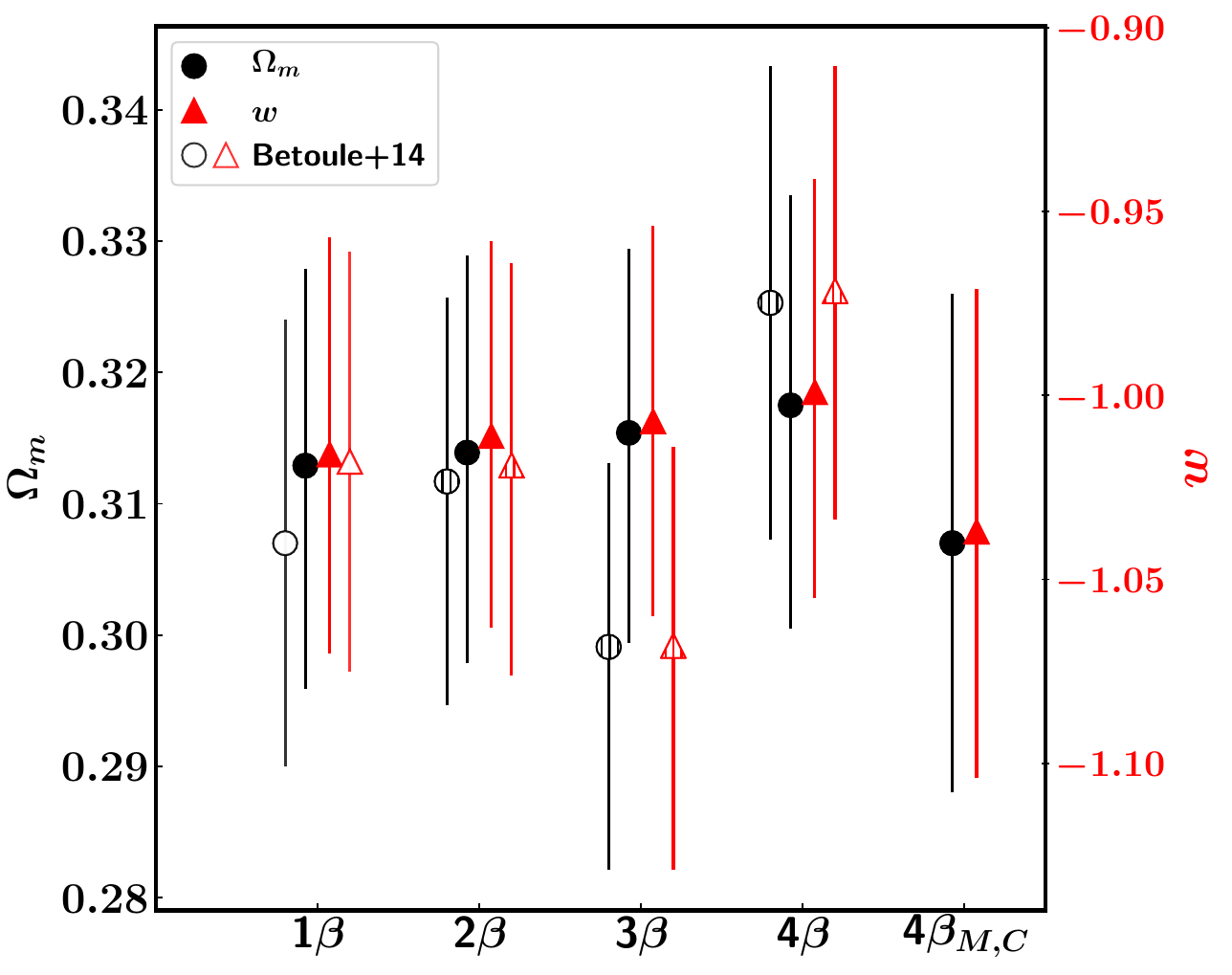}
\caption{Cosmological parameters, $\Omega_m$ (black circles, left ordinate axis) and $w$ (red triangles, right ordinate axis), derived using SNe~Ia + \textit{Planck} for different cosmological fits allowing several $\beta$ parameters when the sample is split according to the host-galaxy stellar mass (full symbols) and to the colour (hatched symbols). 4\,$\beta_{M,C}$ corresponds to the split based on two host-galaxy stellar mass bins and two colour bins. Black and red empty symbols represent the values obtained by \citet{Betoule14}.}
\label{fig:beta_step_cosmo}
\end{figure}

We first investigate the impact on cosmology if several $\beta$ populations are allowed in the fit. In Figure \ref{fig:beta_step_cosmo}, changes in the dark energy equation-of-state parameter ($w$) and the matter density ($\Omega_{M}$) are shown for the cases in which different $\beta$ values are allowed, when the data are split according to the host-galaxy stellar mass and/or colour. When divided into two host-mass bins, for example, the first bin corresponds to SNe in low stellar mass galaxies and the second bin to SNe in high stellar mass galaxies, and each of these subgroups has a $\beta$ parameter inferred from the fit. As can be seen, the cosmological parameters are consistent with the standard 1 $\beta$ fit, and no significant variations are visible when using 2, 3, or 4 bins. If we split the data with respect to other observables ($m_B$, $z$, $X$, and $\mathcal{C}$), the results are similar; H$_0$, $w$, and $\Omega_{M}$ are consistent within their uncertainties independent of the number of bins. We also show the comparison of our cosmological parameters to those derived by \citet{Betoule14} for a single $\beta$.

\subsection{$\beta$ variation with host-galaxy stellar mass}\label{split_mass}

Although there is no significant change in cosmology with the current data precision and our methodology, we do find important differences between the recovered $\beta$ parameters, as well as clear preferences (based on BIC and Hubble residuals) for some of the fits with multiple colour-luminosity relations. In particular, when the sample is split by redshift, light-curve width, or peak magnitude, no statistical differences are seen between the fitted 1, 2, 3, or 4 $\beta$ values. However, when we split according to the host-galaxy stellar mass or SN colour, significant differences between the various $\beta$ parameters are found. 

\begin{figure*}
\centering
\includegraphics[width=1.0\columnwidth]{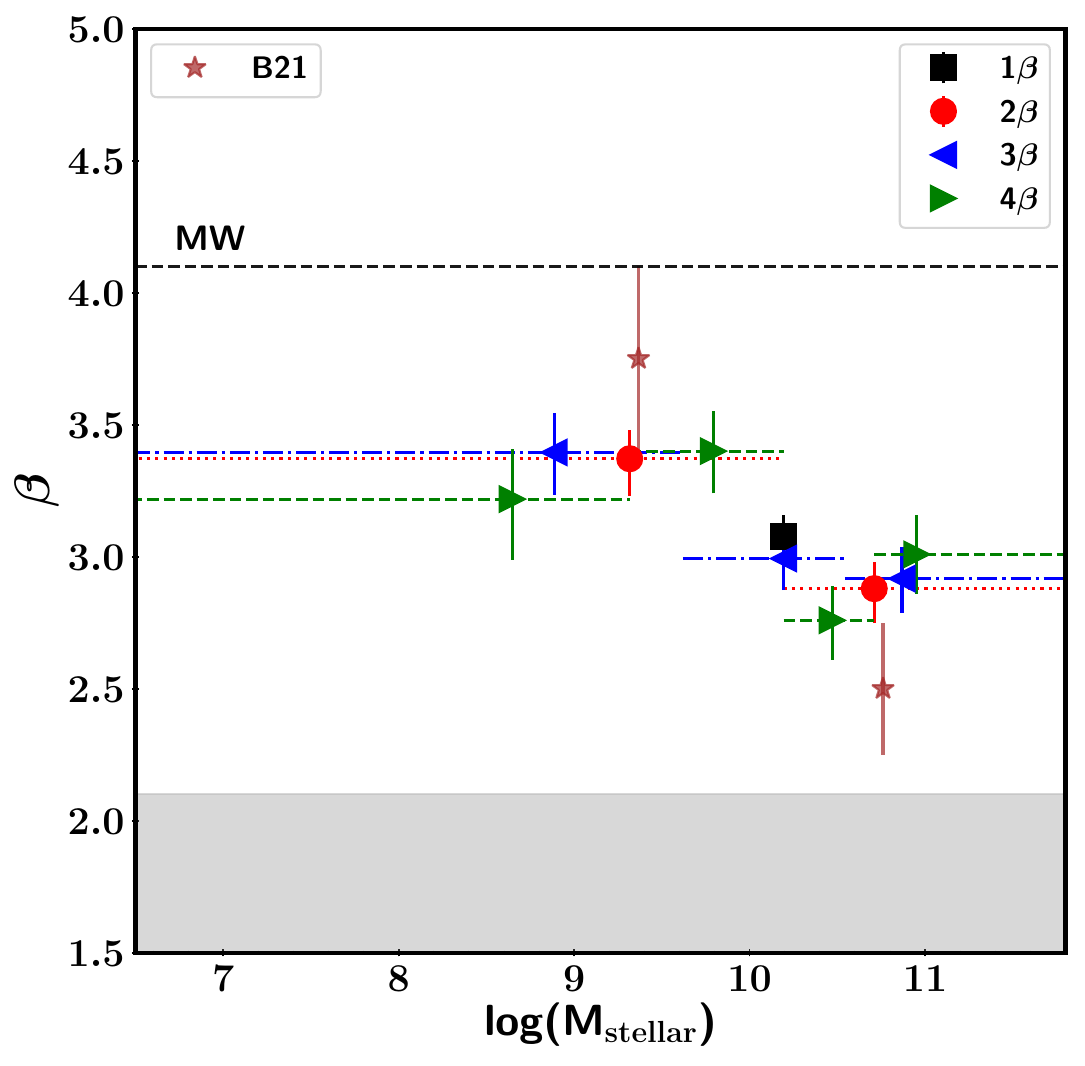}\includegraphics[width=1.0\columnwidth]{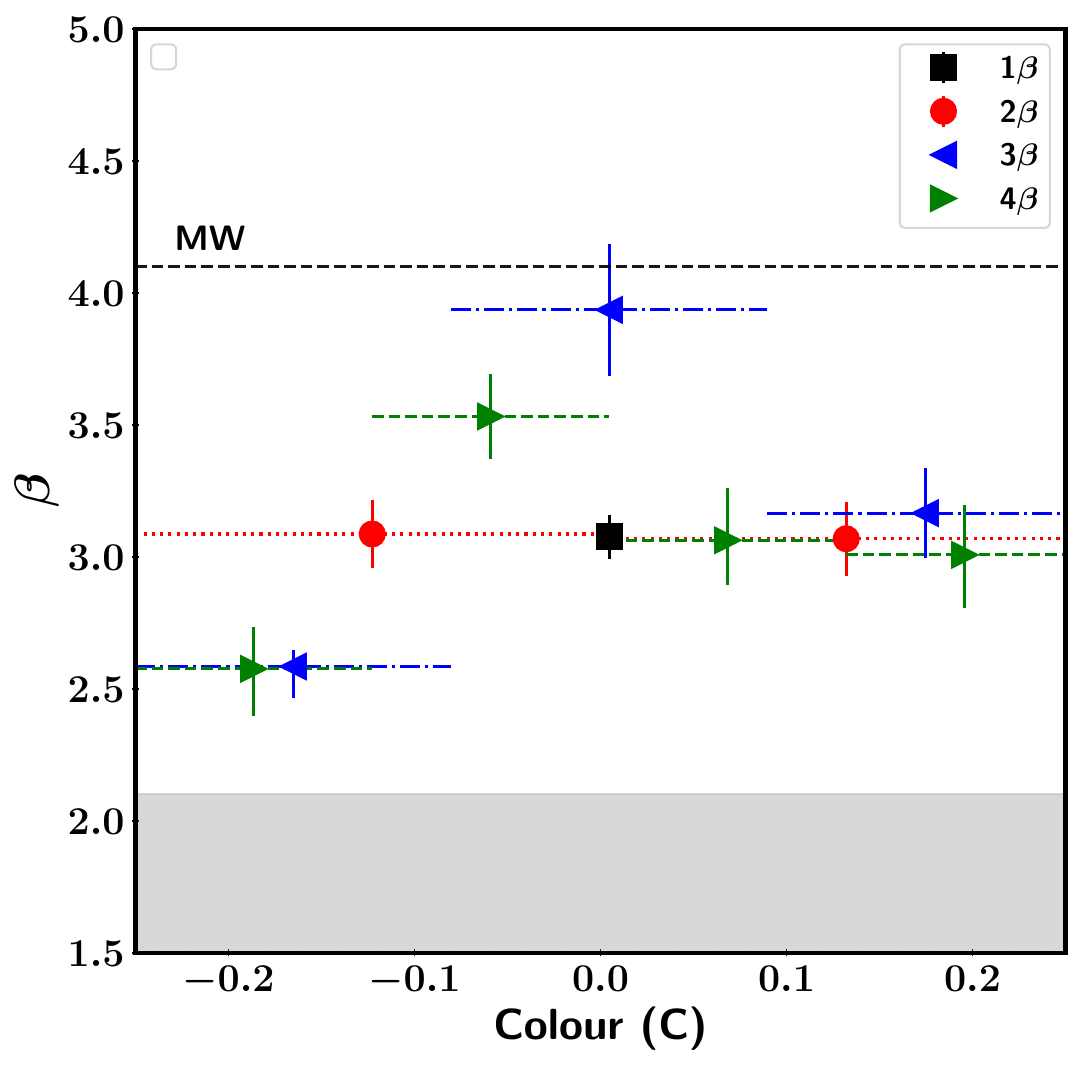}
\caption{\textit{Left:} $\beta$ versus host-galaxy stellar mass. Black square, red circles, blue left-pointing triangles, and green right-pointing triangles are the values derived when the SN~Ia sample is split into 1, 2, 3, and 4 bins (respectively) with respect to the host-galaxy stellar masses. The error bars in $x$ represent the bin width.
Brown stars are from Figure 6b of \citet{Brout20}. 
The horizontal black dashed line represents the average Milky Way $R_{V}=3.1$ ($\beta=4.1$) and the grey filled region is the mean intrinsic colour-luminosity correlation coefficient found by \citet{Brout20}: $\beta_{\mathrm{int}}=1.8\pm0.3$. \textit{Right:} The same as for the left panel but the sample is split according to the observed SN colour at maximum brightness.}
\label{fig:beta_step}
\end{figure*}

Figure~\ref{fig:beta_step} (left panel) presents $\beta$ versus host-galaxy stellar mass obtained from the fits with 1, 2, 3, and 4 mass bins adopted ensuring the same number of SNe per bin (but different width). The $\beta$ values differ by up to $\sim3.1\sigma$ (see Table~\ref{tab:beta}). We obtain similar differences when we bin the data based on the same width (but different number of SNe per bin). 
In this case, we see a step function where higher $\beta$ values of $\sim 3.4\pm0.15$ are favoured at low galaxy stellar mass, $\log(M_{\mathrm{stellar}}/M_{\sun}) < 10$, while smaller $\beta$ values of $\sim 2.9\pm0.12$ are seen at high host galaxy stellar mass, $\log(M_{\mathrm{stellar}}/M_{\sun}) > 10$.  These differences of 2.5--3$\sigma$ in $\beta$ can be explained if there is a difference in the reddening law of galaxies according to their mass, as found in the previous section and as shown by \citet{Salim18}. Our results are also consistent with \citet[][brown stars in Fig. \ref{fig:beta_step}]{Brout20}, where their best-scatter models fit the binned Hubble diagram residuals with $R_V=2.75\pm 0.35$ ($\beta=3.75$) for $\log(M_{\mathrm{stellar}}/M_{\sun}) < 10$ and $R_V=1.5 \pm 0.25$ ($\beta=2.5$) for $\log(M_{\mathrm{stellar}}/M_{\sun}) > 10$. We note that despite the differences in $\beta$, the cosmological fit with a single $\beta$ parameter is still favoured over the 2, 3, and 4$\beta$ according to their BIC values and Hubble residuals (see Table~\ref{tab:beta}).

\begin{figure}
\centering
\includegraphics[width=1.0\columnwidth]{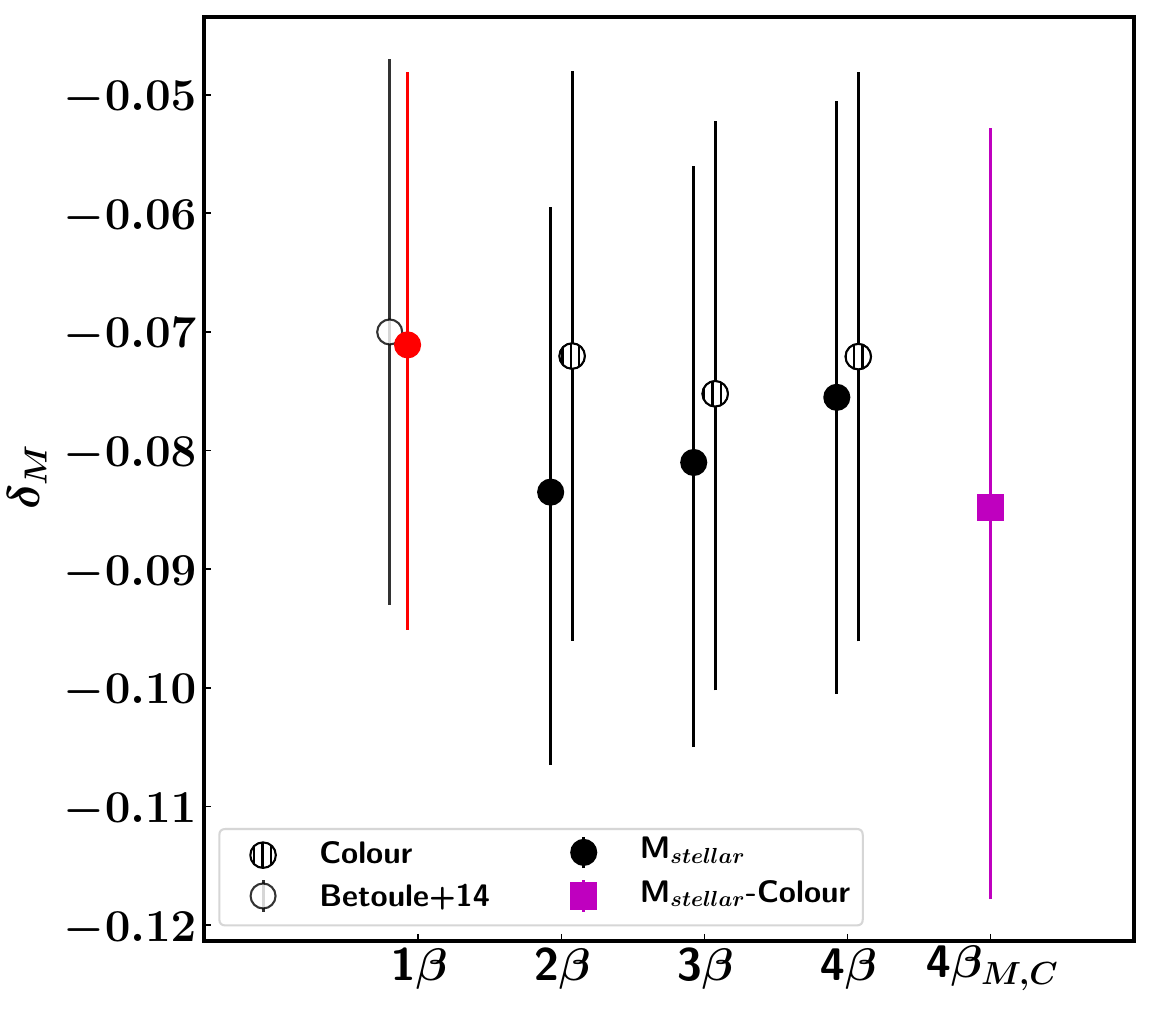}
\caption{Mass-step $\delta_M$ for cosmological fits where 1, 2, 3, and 4 $\beta$ parameters are left free in the fit, separating them according to host-galaxy stellar mass (filled black circles), colour (hatched black circles), and a combined mass \emph{and} colour (purple square). The mass-step correction does not change significantly and is always consistent with a value smaller than zero (by 2--4$\sigma$). Our 1 $\beta$ value is shown in red and the value derived by \citet{Betoule14} as an empty black circle.}
\label{fig:stepmass}
\end{figure}

The found $\beta$ behaviour with $M_{\mathrm{stellar}}$, the $\beta$-$M_{\mathrm{stellar}}$ relation, is not continuous and is strikingly reminiscent of the mass-step used to reduce the Hubble residuals. It happens at a similar division point of around $10^{10}\,M_{\sun}$ \citep[e.g.,][]{Sullivan10}. 
So, this naturally raises the question of whether the usual Hubble residual mass-step is indeed related to this $\beta$-step. However, Figure~\ref{fig:stepmass} reveals that the fitted mass-step is always present with a significance of more than $2.5\sigma$, regardless of the multiple $\beta$ splits we use. Furthermore, when the mass-step term is removed from the cosmological fits with various $\beta$ parameters, the BIC and Hubble residuals are always worse than for the respective cases with $\delta_M$.
Unlike \citet{Brout20}, this argues for a mass-step to optimise the calibration of SNe~Ia, even when there is a $\beta$-step relation with galaxy mass. This is consistent with recent findings of a mass-step even at longer wavelengths, where extinction effects are less important \citep{Ponder20,Uddin20}.

Nonetheless, if we remove the mass-step term in the cosmological fits, the differences between the $\beta$ parameters split according to galaxy mass decrease and are much less significant (1--2$\sigma$). So, there does seem to be a relation between the mass-step and the $\beta$-step after all: SNe in high-mass galaxies have both a different colour-luminosity relation than in low-mass galaxies \emph{and} an additional constant term, the mass-step, $\delta_M$. Since these two terms are related, it is possible that the constant term actually stems from a different intrinsic colour being absorbed in the ``mass-step''; for SNe in high-mass galaxies (see Eq.~\ref{eq_standardisation_int}): $\beta\mathcal{C}+\delta_M\approx (R_V+1) E(B-V)+\beta_{\mathrm{int}}c_{\mathrm{int}}$. Within a SN subgroup, it is more likely that the intrinsic parameters are constant while the dust absorption, $E(B-V)$, changes from SN to SN depending on the dust content. So, the constant mass-step is presumably related to the different intrinsic behaviour of a SN population in a certain galaxy mass range, $\delta_M \sim \beta_{\mathrm{int}}c_{\mathrm{int}}$, while the $\beta$-step is related to the reddening, $\beta\mathcal{C} \sim (R_V+1)E(B-V)$.

These are obviously simplifications since we know that the observed colour is the sum of the colour excess and intrinsic colour, $\mathcal{C}=c_{\mathrm{int}}+E(B-V)$, which are not separated in our analysis, but it explains why when the intrinsic colour is not correctly accounted for (the mass-step), then the colour excess is also incorrect and the dust-reddening relation is affected (the $\beta$-step). This interpretation suggests that both intrinsic and extrinsic dust properties of SNe correlate with galaxy mass, and both in the same direction, making objects of a given colour fainter in lower-mass than in higher-mass environments. Or, in other words, SNe of the same observed luminosity (at the same distance) are bluer in low-mass galaxies and redder in high-mass galaxies.

The above trends provide a coherent picture that agrees with the observations of different SN populations according to galaxy properties: high-velocity SNe that are also intrinsically redder \citep{Mandel14,Foley11b}, with a different inner asymmetry \citep{Cartier11,Maeda10}, happen in more luminous, massive hosts \citep{Pan20,Wang13}, but at the same time they suffer from more dust extinction \citep{Forster12} with variations in the $R_V$ reddening \citep{Forster13}.

\begin{figure}
\centering
\includegraphics[width=1.0\columnwidth]{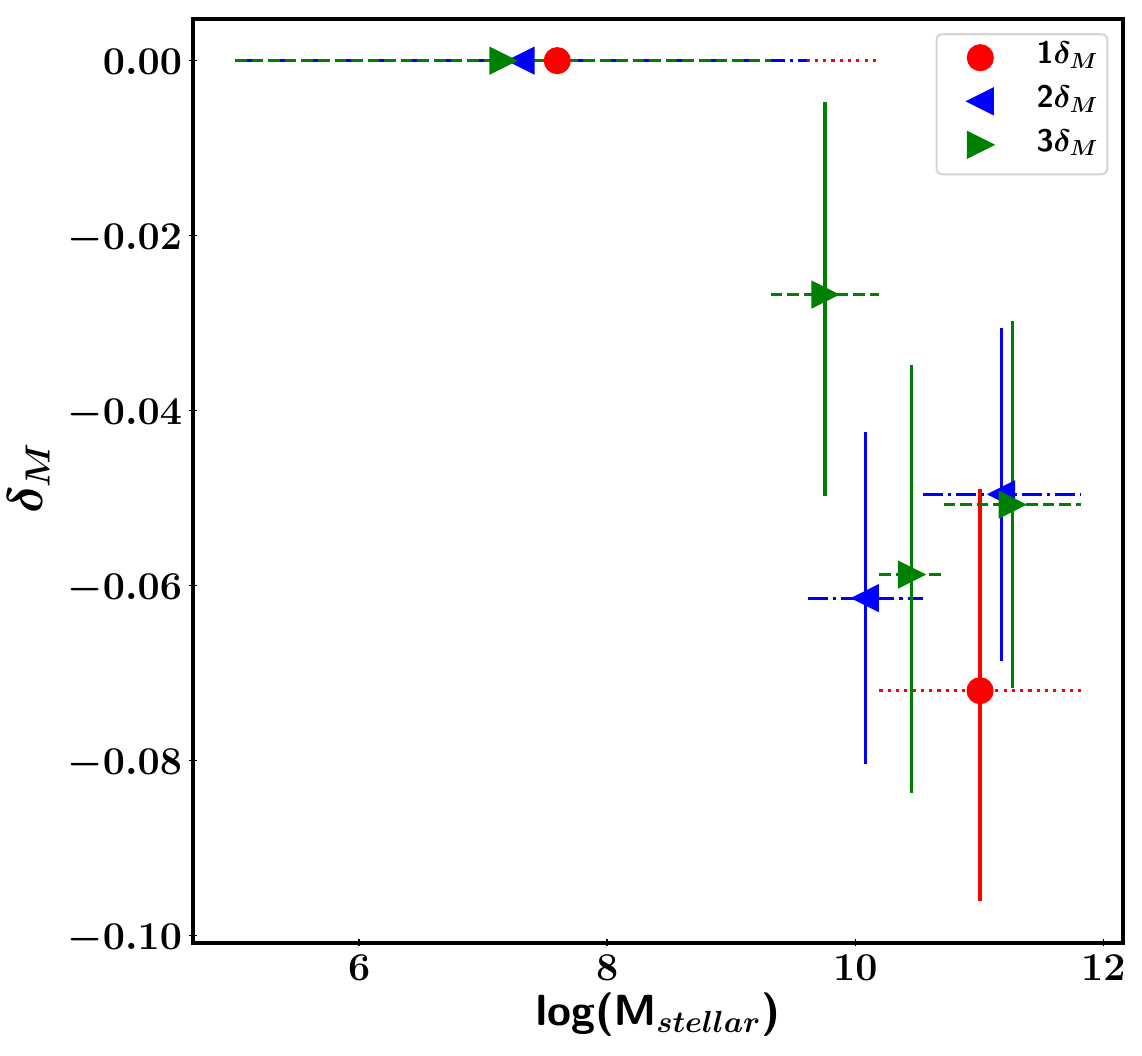}
\caption{$\delta_{M}$ versus host-galaxy stellar mass for different binning of the SN~Ia sample according to galaxy mass: 2, 3, and 4 bins shown in red circles, blue left-pointing triangles, and green right-pointing triangles, respectively. For each split, we fix the first bin to have $\delta_{M} = 0$ and attribute to the other $N-1$ bins different $\delta_{M}$ values. The error bars in $x$ represent the bin length. 
}
\label{fig:step}
\end{figure}

Since the $\beta$-step is strongest when the mass-step is present, the exact division point in mass might just be an artifact of the binning in the fit. We investigate this by fitting several mass-steps. The result is shown in Figure~\ref{fig:step}, where we see that, to the extent of the SN sample and splitting used, a bimodal mass-step is indeed preferred. This is true regardless of whether we use single or multiple $\beta$ values assigned to each mass-step bin.

To summarise, the SN dataset always requires the mass-step  correction in order to increase the precision of the distance calibration. When this mass-step is removed, the found $\beta$-step (i.e., the change in $\beta$ for SNe in different host-galaxy mass ranges) decreases, arguing that the two may be related. The most straightforward explanation is that the mass-step reveals a different intrinsic colour-luminosity component for SNe~Ia in low-mass and high-mass galaxies, while the $\beta$-step tells us there is a difference in the reddening law of these two types of galaxies.

\subsection{$\beta$ variation with SN colour}\label{split_color}

Similarly to galaxy mass, we find significant differences (of up to $5.3\sigma$) for the $\beta$ values obtained when splitting the sample in SN colour intervals. Figure~\ref{fig:beta_step} (right panel) depicts the $\beta$ parameters versus colour for 1, 2, 3, and 4 colour bins divided according to the bin width; each bin has the same width but a different number of objects. All the results are also presented in Table~\ref{tab:beta}.

In this figure, we clearly see that $\beta$ changes with observed SN colour. First, for bluer colours, $\beta$ is small ($\sim 2.5\pm0.2$), then at redder colours it becomes larger reaching a maximum ($\sim 3.9\pm0.3$), and finally at even redder colours it appears to decrease again ($\sim 3.0\pm0.2$), with differences of 3--5$\sigma$.  Interestingly, when split into only two bins (red points), one may be led to think erroneously that there is no variation with SN colour \citep{Sullivan11}. This seemingly puzzling behaviour of $\beta$ with SN colour can be nonetheless explained in light of recent studies. On the one hand, we know that the $\beta$ approach we follow (Eq.~\ref{eq_standardisation}) is a simplification of a more complete treatment in which both intrinsic and extrinsic colour-magnitude components (Eq.~\ref{eq_standardisation_int}) play a role. Both \citet{Brout20} and \citet{Mandel17} find that the intrinsic SN colour-luminosity relation ($\beta_{\mathrm{int}}$) has low values of 1.8--2.2, which should dominate at blue colours in the absence of host extinction. This is represented with the grey shaded area in Figure~\ref{fig:beta_step}. On the other hand, as the colour excess from reddening along the line of sight becomes more important at redder colours, $R_V$ plays an increasingly dominant role, so that the maximum value gets closer to the average extinction law of the MW ($R_V=3.1$; $\beta=4.1$). Since our standard \citet{Tripp98} approach combines both components into a single one (even if we divide $\beta$ into several bins), the obtained $\beta$ value at blue colours is higher than the actual $\beta_{\mathrm{int}}$, and the maximum measured $\beta$ at redder colours is smaller than the one from the MW. This is also predicted by \citet[][see their Fig. 3]{Mandel17}.

Moreover, the decrease of $\beta$ at the reddest SN colours could be best explained by the decrease in $R_V$ at higher extinctions, a trend that has been observed
along the line of sight to SNe~Ia: the average $R_V$ seems to decrease with increasing $E(B-V)$, as shown in Figure~19 of \citet{Burns14} and Figure~8 of \citet{Mandel11}.

We find that the BIC favours all models with 2, 3, and 4 $\beta$ values, although the Hubble residuals are lower only for the latter. Also, when we remove the mass-step from the fits, the differences in $\beta$ are still significant for all the splits (2, 3, and 4 $\beta$), showing that this relation is probably more fundamental than for the galaxy-mass case: all SNe have both intrinsic and extrinsic colour-luminosity relations, regardless of their environment. It is important to note that given our likelihood (Eq.~\ref{eq:likelihood}), the fitted $\beta$ for a SN sample will be affected by the sample distribution of the colour and the colour uncertainty, a possible bias that will be further explored in Section~\ref{sec:like}.

To summarise this section, we find that the $\beta$ evolution with colour is compatible with a low intrinsic colour-luminosity relation that manifests itself better at bluer colours and an external reddening law that is more apparent at redder colours. We claim that the decrease of the $\beta$ parameter at the reddest colours corresponds to a decrease in the reddening law at highest dust absorption (although see Sec.~\ref{sec:like}). 
In the next section, we attempt to merge both dependencies of $\beta$: galaxy mass and SN colour. 

\subsection{$\beta$ variation with SN colour and host-galaxy mass}\label{split_colmass}

\begin{figure}
\centering
\includegraphics[width=1.0\columnwidth]{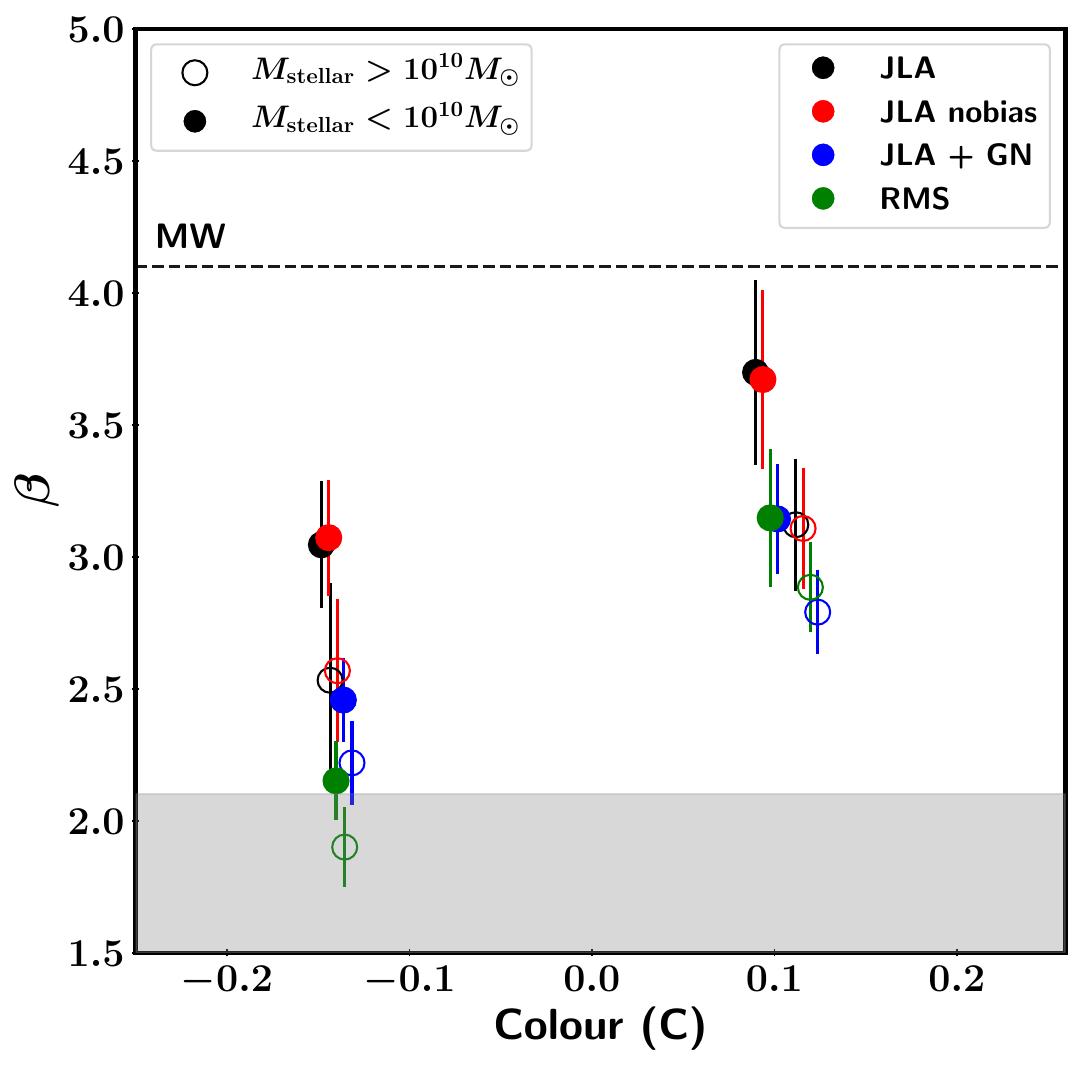}
\caption{The obtained $\beta$ values when divided according to observed colour \emph{and} host-galaxy stellar mass. Empty (filled) symbols represent the SNe~Ia in high (low) host-galaxy stellar mass ($>10^{10}\,M_{\sun}$). The symbol colours represent different likelihoods (see Sec.~\ref{sec:like}: standard JLA (black), JLA without bias correction (blue), JLA with Gaussian normalisation (red) and a RMS minimisation (green).  The horizontal blue dashed line represents the average Milky Way $R_{V}=3.1$ and the grey filled region is the intrinsic colour-luminosity correlation coefficient found by \citet{Brout20}: $\beta_{\mathrm{int}}=1.8\pm0.3$.}
\label{fig:beta_colmass}
\end{figure}

Given the $\beta$ dependence we find with SN colour and host-galaxy mass, here we combine the two effects. The SN sample is split into four bins of equal number of objects according to host-galaxy stellar mass \emph{and} colour. First, we split into two bins with respect to host-galaxy stellar mass, and then we split those two bins into two other bins based on SN colour (with the inclusion of the usual mass-step). The results can be seen in Figure~\ref{fig:beta_colmass} and Table~\ref{tab:beta}. Consistent with our previous results, bluer SNe~Ia for each galaxy stellar mass bin favour smaller $\beta$ values, while redder SNe~Ia have larger $\beta$. Conversely, from the point of view of the galaxy mass, we confirm that larger colour-luminosity relations are preferred in less-massive hosts while smaller colour-luminosity relations are favoured in more-massive galaxies. However, this is statistically less significant than when dividing only according to mass (see Section~\ref{split_mass}), and it is even less significant for blue SNe~Ia. This can be explained as the intrinsic colour-luminosity relation, $\beta_{\mathrm{int}}$, is dominant for bluer SNe~Ia, while for redder SNe~Ia the $\beta$ value follows more closely the dust $R_V$ component. This would confirm that the $\beta$--$M_{\mathrm{stellar}}$ relation is related to host extinction, as discussed in the previous section, with different reddening laws in galaxies of different mass \citep{Brout20,Salim18}. The mass-step is again required with $2.7\sigma$ significance. The combined 4 $\beta$ model provides the largest improvement in root-mean square (RMS) of the Hubble residuals even if its associated BIC tells us that it is a fit less favoured than the standard 1 $\beta$ cosmology (see Table \ref{tab:beta}). The reason for this may be related to a biased likelihood, as further discussed in Section~\ref{sec:like}. We note that if we attempt to have two mass bins and three colour bins, this 6 $\beta$ fit does not converge according to the Gelman-Rubin criterion.

Conforming to the obtained intrinsic dispersion ($\sigma_{\mathrm{int}}$) for the different subpopulations  (Table~\ref{tab:beta}), bluer SNe in low-mass galaxies are the most homogeneous sample with an intrinsic dispersion of 0.07. 
Red SNe are clearly disfavoured for precise cosmology, as also suggested by \citet{Kelsey20}. On the one hand, there is a strong extinction contribution that makes red, extinguished SNe worse standard candles. But on the other hand, this can also be explained through the intrinsic colour: it has been recognised in the past that low-velocity, blue SNe are better distance beacons \citep{Foley11b,Siebert20}. Simultaneously, \citet{Kim18} find that low-mass environments offer a cleaner population of calibrators, which is presumably the result of less scatter in local star-forming environments \citep{Rigault13}. Given that intrinsically bluer SNe do occur in low-mass galaxies \citep{Pan20}, less luminosity scatter for blue SN colours and low-mass environments stems partly from the same effect. Thus, red SNe probably have larger luminosity scatter from a mixture of both a different intrinsic colour population and a larger extinction, not easily corrected with a simple linear colour-luminosity relation. We thus propose that a combination of blue and low-mass environments might offer yet the best set of standard candles.

\section{Likelihood systematics}\label{sec:like}

The likelihood and methodology described in Section~\ref{txt:montepython} has been widely used in the literature. However, it is known that it has its shortcomings and may introduce biases in the obtained nuisance parameters, notably $\alpha$ and $\beta$ \citep[e.g.,][]{Kessler17}. 

On the one hand, an important problem arises from the fact that the width of the parent distribution of SN parameters such as the colour and light-curve width are similar to their uncertainties. A proper methodology should therefore attempt to model the inherent distributions of the parameters and their uncertainties, as well as the different physical components of the SN~Ia model such as the intrinsic colour and dust extinction separately. This needs to be done with care, since wrong assumptions can have important biases in cosmological fits, as previously shown \citep{Kessler09,Wood-Vasey07, Conley07}. Although computationally expensive, full hierarchical Bayesian treatments deal with some of these issues \citep{Mandel20,Hinton19,Rubin15,March11}. Aside from this, in our current approach, the covariances $\mathsf{C}$ (Eq.~\ref{eq:sigma}) depend on the nuisance parameters ($\alpha$, $\beta$), therefore introducing biases in them. In an attempt to diminish such biases, the likelihood of Equation~\ref{eq:likelihood} has been commonly trimmed of an additional term, the Gaussian normalisation $\log(\det\mathsf{C})$ \citep[][Appendix B]{Conley11}.

On the other hand, the bias corrections should account for selection effects from multiple observables, not just the redshift but also colour and light-curve width. Moreover, the impact of this bias should be evaluated on the fly during the fitting process. These concerns have been addressed with large simulations and alternative cosmological fitting techniques \citep{Kessler17,Kessler19}. 

Another important piece of the cosmological inference resides in the intrinsic dispersion $\sigma_{\mathrm{int}}$ of an SN sample (see, e.g., \citealt{Kessler17,Betoule14}), which also enters the uncertainty budget (Eq.~\ref{eq:sigma}) and is therefore cumbersome to obtain as a free parameter in the fit. Alternatives have relied on finding the $\sigma_{\mathrm{int}}$ value that nears the $\chi^2$ per degree of freedom to 1 for a given cosmology, in an iterative process until reaching a desired convergence (typically after 2-3 iterations, e.g. \citealt{Conley11}) or using the restricted log-likelihood method of \citet[][Eq. 14]{Betoule14}. We have used here the standard $\chi^2$ approach in which each population gets its own intrinsic dispersion, as shown in Table~\ref{tab:beta}.

Powerful alternatives to the standard likelihood have been recently implemented in the study of the reddening laws of different host stellar masses and their effect on the mass-step \citep{Thorp21,Brout20}, and are beyond the scope of this paper. However, we do investigate here simple modifications to the standard likelihood to test for systematic effects in our results. We are mostly interested in the relative change of the colour-luminosity relation ($\beta$) with host-galaxy mass and SN colour, rather than its absolute value. According to \citet[][Table 6]{Kessler17}, a likelihood without bias corrections (and without the Gaussian normalisation) results in a smaller offset in $\beta$. We reconduct our cosmological fits with varying $\beta$ parameters for this likelihood with no bias (``JLA+no bias'') as well as for a likelihood including the Gaussian normalisation that is normally left out (``JLA+GN''). With this procedure, we may also infer directly $\sigma_{\mathrm{int}}$ for each subpopulation as a free parameter (similarly to the restricted log-likelihood method of \citealt{Betoule14}).  Finally, we also maximise a likelihood of just the RMS of the Hubble residuals without the uncertainties and covariances in the parameters (``RMS''). Although this latter methodology neglects important observed weighting of the data, it circumvents the issues of the dependence of covariances with the fitted parameters and the width of the intrinsic dispersion. 

Even though the absolute values change, for these three additional likelihoods we recover the observed trend of higher/lower $\beta$ for low/high-mass galaxies. The significance of this difference for the RMS minimisation is lower at 1.6$\sigma$. The trend with colour is also clearly seen: bluer/redder SNe~Ia have lower/higher $\beta$ at high significance ($>3\sigma$) for all cases. However, we do not see a decrease in $\beta$ at the reddest colours for the JLA+GN likelihood, nor for the RMS minimisation. Additionally, we show in Figure~\ref{fig:beta_colmass} and Table~\ref{tab:likelihoods} the result of the 4$\beta$ split in both host mass and SN colour for the four likelihoods. The relative behaviour is consistent for each likelihood even if the JLA+GN and RMS likelihoods indeed present negative biases with respect to the standard JLA likelihood. Conversely, the dispersion in the Hubble residuals obtained for these two likelihoods is always lower than the standard one, providing evidence for the biased nature of the more standard likelihoods. We therefore caution that the residual dispersion and BIC values should be treated with care. Finally, it is important to note that for all $\beta$ splits of all the likelihoods we study, we always find strong evidence for the presence of the mass-step (see Table~\ref{tab:likelihoods}).

We have shown here that the choice of likelihood (including bias correction and the intrinsic dispersion) may affect the absolute values of the colour-luminosity relation and may not provide the best minimisation of the Hubble residuals; nonetheless, the relative change of $\beta$ according to the SN host stellar mass or colour population is preserved, as well as the necessity of a mass-step.

\section{Conclusions}
\label{sec:conc}
 
By investigating multiple luminosity-colour relations in SN~Ia cosmological fits, we present new insights and possible explanations for several open questions of SN~Ia science: from the optimisation of their standardisation for cosmology, to the origin of the mass-step through the existence of intrinsic SN populations that correlate with the environment, to the presence of multiple reddening laws in different galaxies.
 
Instead of using a single universal colour-luminosity relation in the \citet{Tripp98} calibration of SNe~Ia (1 $\beta$), as is usually done, we have explored the cosmological and physical consequences of assuming various $\beta$ relations for different SNe. To do this, we use a common SN~Ia sample for cosmology \citep{Betoule14}, and follow two simple approaches: (a) we generate plausible simulations of $\beta$ distributions for the observed SNe motivated by informed empirical trends of different SN populations, together with cosmological fits compared to the standard 1 $\beta$ scenario; and (b) cosmological fits with multiple $\beta$ parameters are left free in the fit and compared to the fiducial 1 $\beta$ case, whereby the $\beta$ values are split according to different SN observables: redshift, colour, light-curve width, magnitude, and host-galaxy mass. Our main findings are as follows.
 
\begin{enumerate}[(i)]
\item We obtain a significant relationship between $\beta$ and host-galaxy mass (Fig.~\ref{fig:beta_step}, left panel) that has a step at $\sim10^{10}\,M_{\sun}$ going from higher $\beta$ values at lower mass to lower $\beta$ at higher mass. This is consistent with a variation of $R_V$ with galaxy mass found by \citet{Salim18} and by \citet{Brout20} for SNe~Ia. It may be explained through a different attenuation law in quiescent galaxies that are more massive.\label{item:beta-mass}

\item A dependence of $\beta$ with SN colour is also found (Fig.~\ref{fig:beta_step}, right panel), presumably confirming the presence of an intrinsic colour-luminosity relation, best perceived at bluer colours as predicted by \citet{Mandel17}, and an extrinsic dust reddening with $R_V$ nearing the average Milky Way value at low extinction and possibly decreasing at high extinction, as in \citet{Burns14} and \citet{Mandel11}. This latter decrease is not seen when using other non-standard likelihoods (see section~\ref{sec:like}).

\item We find no evidence for a riddance of the mass-step in any of our cosmological fits with multiple free $\beta$ parameters. This suggests that the $\beta$--$M_{\mathrm{stellar}}$ relation we find in item~(\ref{item:beta-mass}) cannot explain the difference in residual SN luminosity with host mass. This is in agreement with other cosmological analyses \citep{Thorp21} and with the observations of a mass-step in the near-infrared where extinction becomes less relevant \citep{Ponder20,Uddin20}.
 
\item Since the $\beta$--$M_{\mathrm{stellar}}$ dependence of item~(\ref{item:beta-mass}) becomes less significant when we take out the mass-step, we conjecture that the origin of this mass-step is a difference in intrinsic colour-luminosity relation ($\sim\beta_{\mathrm{int}}\mathcal{C}_{\mathrm{int}}$) of two SN populations found in galaxies with different masses: for a given luminosity, SNe~Ia that are intrinsically redder tend to occur in higher-mass galaxies. This is consistent with previous observations of multiple intrinsic colour populations \citep{Mandel14,Foley11,Maeda10} observed in different galaxies \citep{Pan20,Wang13}. Furthermore, given that different $\beta$ values are found for these two groups in item~(\ref{item:beta-mass}), there is \emph{also} a different dust-reddening law for these two types of populations \citep{Forster13}.

 
\item We confirm that SNe that are bluer are better standard candles \citep{Foley11b,Kelsey20,Siebert20}, particularly those in low-mass environments \citep{Kim18,Rigault13}. 
The population of SNe that are blue \emph{and} have less-massive hosts should provide the best distance precision.

\item The cosmological parameters $\Omega_M$ and $w$ do not change within the errors for the studied dataset and for all simulations and multiple $\beta$-free fits we perform. Variations of $|\Delta w| = 0.04$ with the fiducial fit, as found by \citet{Brout20}, are seen for various of our scenarios, but they are completely consistent with each other within the uncertainties.
  
\item Based on the BIC diagnostic and improvement in the Hubble residuals, we find significant evidence for a preference in cosmological fits that contain multiple colour-luminosity relations split according to different SN colours (3 $\beta$) or host-galaxy masses (2 $\beta$), or colour and galaxy mass combined (4 $\beta$). The fits should contain a mass-step correction as well.

Since the accuracy of the fitted cosmology remains unchanged, this is a straightforward addition to routine cosmological analyses that can be implemented right away in current and future datasets. Without loss of predictive power in cosmology, there is a clear gain in the understanding of intrinsic SN properties and extrinsic dust effects.  Furthermore, as the observations of SNe~Ia keep improving and the statistical errors are further reduced, the systematic biases in colour and the colour-luminosity relation keep emerging as fundamental ones, thus requiring a more proper and comprehensive treatment of intrinsic/extrinsic colours and multiple colour-luminosity relations. If we really want to step into a new era of SN cosmology, stringent independent, multitechnique, multiwavelength studies of the colours of individual or grouped SNe~Ia, and the dust in their environments, are needed.
\end{enumerate}

\begin{table*}
\centering
\caption{Cosmological fits with varying colour-luminosity relations.}
\scalebox{0.85}{
\begin{tabular}{c|cccc|cccc|cc}
\hline
\hline
\\[-1em]
$\pmb{1 \beta}$ & $\beta_0$ & & & & $\sigma_0$ & & & & $\Delta\mathrm{BIC}$ & $\Delta \mathrm{RMS}(10^{-2})$\\
\\[-1em]
\hline
\\[-1em]
Mstellar & $3.08 \pm 0.08$ & & & & $0.106 \pm 0.005$ & & & & 0 & 0\\
Colour & $3.08 \pm 0.08$ & & & &$0.106 \pm 0.005$ & & & & 0 & 0 \\
\\[-1em]
\hline
\hline
\\[-1em]
$\pmb{2 \beta}$ & $\beta_1$ & $\beta_2$  & & & $\sigma_1$ & $\sigma_2$ & & & $\Delta\mathrm{BIC}$ & $\Delta \mathrm{RMS}(10^{-2})$ \\
\hline
\\[-1em]
Mstellar & $3.37_{-0.14}^{+0.12}$ & $2.88_{-0.11}^{+0.11}$ & & & $0.105_{-0.009}^{+0.008}$ & $0.106\pm0.007$ & & & 3.31 & 0.038 \\
Colour & $3.09_{-0.13}^{+0.14}$ & $3.07_{-0.13}^{+0.14}$ & & & $0.096_{-0.007}^{+0.006}$ & $0.127\pm0.010$ & & & -10.3 & 0.004 \\ 
\\[-1em]
\hline
\hline
\\[-1em]
$\pmb{3 \beta}$ & $\beta_1$ & $\beta_2$ & $\beta_3$ & & $\sigma_1$ & $\sigma_2$ & $\sigma_3$ & & $\Delta\mathrm{BIC}$ & $\Delta \mathrm{RMS}(10^{-2})$\\
\\[-1em]
\hline
\\[-1em]
Mstellar & $3.40_{-0.16}^{+0.15}$ & $2.99_{-0.12}^{+0.12}$ & $2.92_{-0.13}^{+0.12}$ & & $0.113_{-0.011}^{+0.010}$ & $0.100_{-0.009}^{+0.008}$ & $0.106_{-0.010}^{+0.009}$ & & 8.06 & 0.111\\
Colour & $2.58_{-0.12}^{+0.06}$ & $3.94_{-0.25}^{+0.25}$ & $3.17_{-0.17}^{+0.17}$ & & $0.101_{-0.012}^{+0.011}$ & $0.103\pm0.006$ & $0.153_{-0.022}^{+0.019}$ & & -28.8 & 0.010\\ 
\\[-1em]
\hline
\hline
\\[-1em]
$\pmb{4 \beta}$ & $\beta_1$ & $\beta_2$ & $\beta_3$ & $\beta_4$ & $\sigma_1$ & $\sigma_2$ & $\sigma_3$ & $\sigma_4$ & $\Delta\mathrm{BIC}$ & $\Delta \mathrm{RMS}(10^{-2})$\\
\\[-1em]
\hline
\\[-1em]
Mstellar & $3.22_{-0.23}^{+0.19}$ & $3.40_{-0.16}^{+0.15}$ & $2.76_{-0.15}^{+0.13}$ & $3.01_{-0.15}^{+0.15}$  & $0.116_{-0.013}^{+0.012}$ & $0.100_{-0.011}^{+0.010}$ & $0.108_{-0.010}^{+0.009}$ & $0.104_{-0.012}^{+0.010}$  & 7.02 & -0.036 \\
Colour & $2.58_{-0.18}^{+0.16}$ & $3.53_{-0.16}^{+0.16}$ & $3.06_{-0.17}^{+0.20}$ & $3.01_{-0.20}^{+0.19}$  & $0.135_{-0.022}^{+0.019}$ & $0.090\pm0.007$ & $0.128_{-0.011}^{+0.010}$ & $0.141_{-0.034}^{+0.025}$  & -28.2 & -0.312 \\
\\[-1em]
\hline
\hline
\\[-1em]
$\pmb{4 \beta_{\mathcal{C},M}}$ & $\beta_{\mathcal{C}_1,M_1}$ & $\beta_{\mathcal{C}_1,M_2}$ & $\beta_{\mathcal{C}_2,M_1}$ & $\beta_{\mathcal{C}_2,M_2}$  & $\sigma_{\mathcal{C}_1,M_1}$ & $\sigma_{\mathcal{C}_1,M_2}$ & $\sigma_{\mathcal{C}_2,M_1}$ & $\sigma_{\mathcal{C}_2,M_2}$  & $\Delta\mathrm{BIC}$ & $\Delta \mathrm{RMS}(10^{-2})$\\
\\[-1em]
\hline
\\[-1em]
Colour/Mstellar & $3.05_{-0.24}^{+0.28}$ & $2.53_{-0.37}^{+0.25}$ & $3.70_{-0.35}^{+0.32}$ & $3.12_{-0.25}^{+0.22}$  & $0.074_{-0.011}^{+0.010}$ & $0.101_{-0.011}^{+0.010}$ & $0.128_{-0.013}^{+0.012}$ & $0.108\pm0.010$  & 7.99 & -0.325\\
\\[-1em]
\hline
\hline

\end{tabular}
}
\tablefoot{Colour-luminosity $\beta$ coefficients and intrinsic scatter $\sigma_{\mathrm{int}}$ for 1, 2, 3, or 4 bins in SN colour or galaxy mass, and for 4 bins of combined colour/mass. $\Delta\mathrm{BIC} = \mathrm{BIC} - \mathrm{BIC}_0$ and $\Delta\mathrm{RMS} = \mathrm{RMS} - \mathrm{RMS}_0$ is the difference in BIC diagnostic and Hubble residual RMS
with respect to the standard 1 $\beta_0$ fit.}
\label{tab:beta}
\end{table*}

\begin{table*}
\centering
\caption{Comparison of likelihoods.}
\begin{tabular}{c|cccccccc}
\hline
\hline
Likelihood  & $\beta_1$ & $\beta_2$ & $\beta_3$ & $\beta_4$ & $\delta_M$ & RMS \\
\hline
\hline
JLA -- $\pmb{1\beta}$  & $3.08\pm0.08$  & -- & -- & -- & $-0.071_{-0.024}^{+0.023}$ (3.1$\sigma$) & 0.1690 \\
JLA+nobias -- $\pmb{1 \beta}$  & $3.07\pm0.08$  & -- & -- & -- & $-0.073_{-0.024}^{+0.024}$ (3.0$\sigma$) & 0.1688 \\
JLA+GN -- $\pmb{1 \beta}$  & $2.60\pm0.07$ & -- & -- & -- & $-0.054_{-0.022}^{+0.023}$ (2.4$\sigma$) & 0.1601 \\
RMS -- $\pmb{1 \beta}$  & $2.45\pm0.07$ & -- & -- & -- & $-0.031_{-0.014}^{+0.013}$ (2.4$\sigma$) & 0.1587 \\
\hline
\hline
JLA -- $\pmb{4 \beta}$  & $3.05_{-0.24}^{+0.28}$  & $3.70_{-0.35}^{+0.32}$ & $2.53_{-0.37}^{+0.25}$ & $3.12_{-0.25}^{+0.22}$ & $-0.085_{-0.033}^{+0.032}$ (2.7$\sigma$) & 0.1657 \\
JLA+nobias -- $\pmb{4 \beta}$ &  $3.07_{-0.22}^{+0.24}$ & $3.67_{-0.34}^{+0.26}$ & $2.57_{-0.27}^{+0.27}$ & $3.11_{-0.23}^{+0.21}$ &  $-0.085_{-0.027}^{+0.027}$ (3.2$\sigma$) & 0.1661 \\
JLA+GN -- $\pmb{4 \beta}$  & $2.46_{-0.16}^{+0.18}$ & $3.14_{-0.21}^{+0.24}$  & $2.22_{-0.16}^{+0.16}$ & $2.79_{-0.16}^{+0.16}$ & $-0.056_{-0.024}^{+0.025}$ (2.2$\sigma$) & 0.1576 \\
RMS -- $\pmb{4 \beta}$  & $2.15_{-0.15}^{+0.15}$ & $3.15_{-0.26}^{+0.23}$  &  $1.90_{-0.15}^{+0.15}$ & $2.88_{-0.17}^{+0.18}$ &  $-0.036_{-0.018}^{+0.018}$ (2.0$\sigma$) & 0.1561 \\
\hline
\hline

\end{tabular}
\tablefoot{Colour-luminosity $\beta$ coefficients, mass-step $\delta_{M}$, and Hubble residual RMS for 1 and 4 bins in both SN colour and galaxy mass, for 4 different likelihoods: JLA, JLA without bias correction, JLA with Gaussian normalisation term, and RMS minimisation.}
\label{tab:likelihoods}
\end{table*}

\section*{Acknowledgements}

We thank the anonymous referee for useful comments that improved the paper. We thank Miguel Zumalacarregui and Carlos Garc\'ia Garc\'ia for their help with {\sc MontePython}. We also thank Dan Scolnic, Dillon Brout, Myriam Rodrigues, Claudia Guti\'errez, Yukei Murakami, Ben Stahl, and Joe Anderson for useful discussions. S.G.G., A.M., and A.P.A. acknowledge support by FCT under Project CRISP PTDC/FIS-AST-31546/2017 and UIDB/00099/2020. Support for A.V.F.'s supernova research group at U.C. Berkeley has been provided by the TABASGO Foundation, Gary and Cynthia Bengier (T.d.J. is a Bengier Postdoctoral Fellow), the Christopher R. Redlich Fund, and the Miller Institute for Basic Research in Science (A.V.F. is a Miller Senior Fellow). We also acknowledge the support of NASA {\it Hubble Space Telescope} grant GO-15640 from the Space Telescope Science Institute, which is operated by the Association of Universities for Research in Astronomy, Inc., under NASA contract NAS 5-26555. L.G. acknowledges financial support from the Spanish Ministry of Science, Innovation and Universities (MICIU) under the 2019 Ram\'on y Cajal program RYC2019-027683 and from the Spanish MICIU project PID2020-115253GA-I00.

\section*{Data availability}
The supernova dataset used in this article is available in \citet{Betoule14}, at \url{https://supernovae.in2p3.fr/sdss_snls_jla}.  

\bibliographystyle{aa}
\bibliography{astro}

\label{lastpage}
\end{document}